\documentclass[article,twocolumn,
amsmath,amssymb,aps,prl]{revtex4}

\usepackage{color}
\usepackage{graphicx}
\usepackage{dcolumn}
\usepackage{braket}
\usepackage{bm}
\usepackage{amsmath,amssymb,amsthm}
\usepackage{lipsum} 
\usepackage{hyperref}
\usepackage{comment}
\usepackage{braket}

\usepackage{soul}

\newcommand{\IOPPAS}{Institute of Physics PAS, Aleja Lotnik\'ow 32/46, 02-668 Warszawa, Poland}

\begin{document}

\author{Nikolaos Giovanoudis$^{1,*}$, Navid Kazemiseresht$^{2,*}$, Fabio Mezzacapo$^1$, Emilia Witkowska$^2$, and Tommaso Roscilde$^1$}
\affiliation{%
$^1$  ENS de Lyon, Universit\'e Lyon 1, CNRS, Laboratoire de Physique, F-69342 Lyon, France
}%
\affiliation{%
$^2$ \IOPPAS
}%
\affiliation{%
$^*$ These authors contributed equally
}%

\title{Universal spin-squeezing dynamics in spinor condensates}

\begin{abstract}
The production of large-scale entangled states is one of the main goals of next-generation quantum technologies, with an immediate potential for applications in the context of entanglement-assisted quantum sensing. A very promising platform to achieve this goal is offered by ultracold spinor gases, made of atoms with a large internal spin sensitive to magnetic fields. Here we show that the native spin-changing collisions in a spinor Bose-Einstein condensate, combined with an arbitrary quadratic Zeeman shift, can generate scalable spin squeezing in the collective spin of the ensemble, following the universal paradigm of the celebrated one-axis-twisting model. Squeezing dynamics is driven by the quadratic Zeeman shift when this shift is small; and by the spin-changing collisions for large shifts, in the form of stroboscopic squeezing. Turning off the Zeeman shift freezes out the collective-spin dynamics, so that the ensuing collective spin dynamics can be uniquely governed by an external field to be sensed.  Our theoretical results pave the way for the use of spinor Bose gases with a large spin in fundamental studies of entanglement, as well as in advanced metrological applications. 
\end{abstract}

\maketitle

\emph{Introduction.} Many-body entanglement, originally recognized as a paradoxical feature of quantum mechanics \cite{Gilder-book}, has surged to the role of essential resource in quantum information and future quantum technologies \cite{IkeandMike,Georgescuetal2014,Pezzeetal2018}. In particular, multipartite entanglement \cite{Horodeckietal2009} implies that several degrees of freedom are involved in a joint superposition state which is not reducible to a product of smaller superpositions; and it represents the fundamental resource for entanglement-assisted quantum sensing \cite{Pezzeetal2018,Frerotetal2023}. 

Different forms of multipartite entangled states have been realized in experiments on state-of-the-art many-body quantum devices, from superconducting circuits \cite{Song2019, javadiabhari2025} to trapped ions \cite{Monzetal2011,Frankeetal2023} and neutral atoms \cite{Esteve2008,Riedel2010,Lueckeetal2011,Gross2012,Luecketal2014,Luoetal2017,Zouetal2018,Bravermanetal2019,Bornetetal2023,Finkelstein2024,Cao2024}. Among them, spin squeezed states \cite{Ma2011} of spin ensembles emerge when entanglement brings the noise of one collective spin component below the value for uncorrelated spins; and they offer an outstanding resource for metrology. Their use to enhance the precision of quantum interferometers has been recently demonstrated \cite{Muessel2014PRL,Pedrozo-Penafiel2020,Schulte2020,Greve2022,Eckneretal2023,Frankeetal2023}. 

Most of the experimental as well as theoretical work to date has focused on ensembles of qubits (or two-level systems); nonetheless, \emph{qudits} (or $d$-level systems with $d>2$) represent a very valuable alternative. Indeed the larger dimension of their local Hilbert space leads to a significant enhancement of the maximal sensitivity of quantum states to unitary transformations, scaling as $d^2$ \cite{Pezzeetal2018}. Several platforms allow for the realization of qudit ensembles \cite{Moreno-Pinedaetal2018,Erhard2020,Blok2021,Qiao2025}. 
Among them, spinor atomic gases \cite{Stamper-Kurn-Ueda_2013,KawaguchiU2012} with a large internal spin $S>1/2$, realizing qudits with $d=2S+1$, stand out for their significant potential, since entanglement between the atomic spins can be induced either by spin-dependent contact interactions \cite{Lueckeetal2011,Luecketal2014,Luoetal2017,Zouetal2018} or dipolar interactions \cite{Chomazetal2022}. 
So far, spin-dependent contact interactions have been shown to lead to spin-nematic squeezing in spinor gases~\cite{Hamley2012,Mao2023}, which involves linear and quadratic operators in the collective-spin components.
On the other hand, spin squeezing of large-$S$ spin ensembles, along with its potential for metrological applications in Ramsey interferometry, remains elusive so far. 
Indeed spin squeezing is typically generated starting from a factorized, coherent spin state \cite{KitagawaU1993}; but this state is an eigenstate of the rotationally invariant spin-dependent contact interactions of spinor gases, which therefore cannot evolve it into an entangled state \cite{Lepoutreetal2018}.  

In this work we theoretically show that scalable spin squeezing can be naturally obtained in spinor Bose-Einstein condensates (BEC) by quadratic Zeeman energy. 
%
In the case of $S=1$ atoms (such as $^{87}$Rb or $^{23}$Na \cite{Stamper-Kurn-Ueda_2013}) the quadratic Zeeman term breaks the spin isotropy of contact interactions, thereby inducing spin-squeezing dynamics that exhibit scaling behavior as in the paradigmatic one-axis-twisting (OAT) model \cite{KitagawaU1993}.
%
The scalability of spin squeezing is universal for all regimes of spin-1 BEC, and hence extremely robust. 
The dynamics of the collective spin can be stopped by turning off the quadratic Zeeman field, making squeezing a stationary property.
Our conclusions are based on mapping of the full Hamiltonian of the system onto effective models, and on exact numerical simulations.

\emph{Model Hamiltonian.}
We consider a spin-1 BEC with $N$ bosonic atoms having $S=1$ and occupying the same spatial mode 
wave function $\phi(r)$~\cite{PhysRevLett.81.5257, PhysRevA.102.023324}. The whole physics of the system is described in terms of $2S+1$ spin modes, and related creation-destruction operators $a_m, a_m^\dagger$ $(m=-S,.., S)$ for atoms in those modes, where $m$ is the value of the $S^z$ spin component for a single atom; and $N=\sum_m a_m^\dagger a_m$ is a constant.
Considering the spin-dependent interaction term and quadratic Zeeman energy, the Hamiltonian of the system reads \cite{Stamper-Kurn-Ueda_2013,KawaguchiU2012}
\begin{equation}
\frac{H}{c} 
= \frac{{\bm J}^2}{2 N}   + q \sum_m m^2 a_m^\dagger a_m 
\label{e.H}
\end{equation}
where ${\bm J} = (J_x, J_y, J_z)$ is the collective spin operator, with $J_\mu = \sum_{m,m'} a_m^\dagger (S_\mu)_{m,m'} a_{m'};$ and $S_\mu$ a spin-$S$ matrix ($\mu=x,y,z$). 
Here $c = N c_2 \int d^3 r |\phi(\bm r)|^4$ is the spin-dependent interaction coefficient, and $c_2$ is defined in terms of s-wave scattering lengths~\cite{KAWAGUCHI2012253}.
A positive (negative) $c$ corresponds to
antiferromagnetic (ferromagnetic) interactions. In what follows all energies are expressed in units of $c$ (including its sign), and times in units of $\hbar/|c|$; moreover we set $\hbar=1$.
%
The parameter $q$ expresses the quadratic Zeeman shift in reduced units, and its sign reflects the relative sign between the actual quadratic shift and the interaction. $q$ can be controlled either by a magnetic field or by light~\cite{Chalopinetal2018, Evrard2019,Niezgoda_2019}. In the case of $S=1$ atoms, to which we shall specialize the following discussion, the above Hamiltonian contains all the relevant terms of spin-changing contact collisions; and the quadratic Zeeman shift term can be conveniently rewritten as $-q N_0 + {\rm const.}$, with $N_0 = a_0^\dagger a_0$. 

In this work we focus on the unitary dynamics $|\psi(t)\rangle = e^{-iHt} |\psi(0)\rangle$ starting from a coherent spin state (CSS) aligned with the $x$ axis, $|\psi(0)\rangle = |{\rm CSS}_x\rangle$. The CSS is such that $J_x |{\rm CSS}_x\rangle = NS |{\rm CSS}_x\rangle$, and for $S=1$ spins, it takes the form $|{\rm CSS}_x\rangle = \frac{1}{2^N\sqrt{N!}} \left ( a_1^\dagger + \sqrt{2} ~a_0^\dagger + a_{-1}^\dagger \right )^N |0\rangle$. It has maximal collective-spin length $\langle {\bm J}^2\rangle = NS(NS+1)$, and it lives in the subspace spanned by Dicke states $|J=N, M\rangle$, i.e. joint eigenstates of ${\bm J}^2$ and $J_z$ with maximal collective-spin length $J$. The CSS is clearly an eigenstate of the spin-dependent interaction in Eq.~\eqref{e.H}, but it can evolve thanks to the quadratic Zeeman shift. This latter term is a one-body term, and one might erroneously conclude that its inclusion cannot lead to spin entanglement. Nonetheless it does not commute with the spin-dependent interaction Hamiltonian: this aspect leads to a highly non-trivial interplay between the two Hamiltonian terms, resulting in the production of spin-spin correlations. One can analytically understand this aspect in two significant limiting cases. 

\begin{center}
\begin{figure*}[ht!]
\includegraphics[width=1\textwidth]{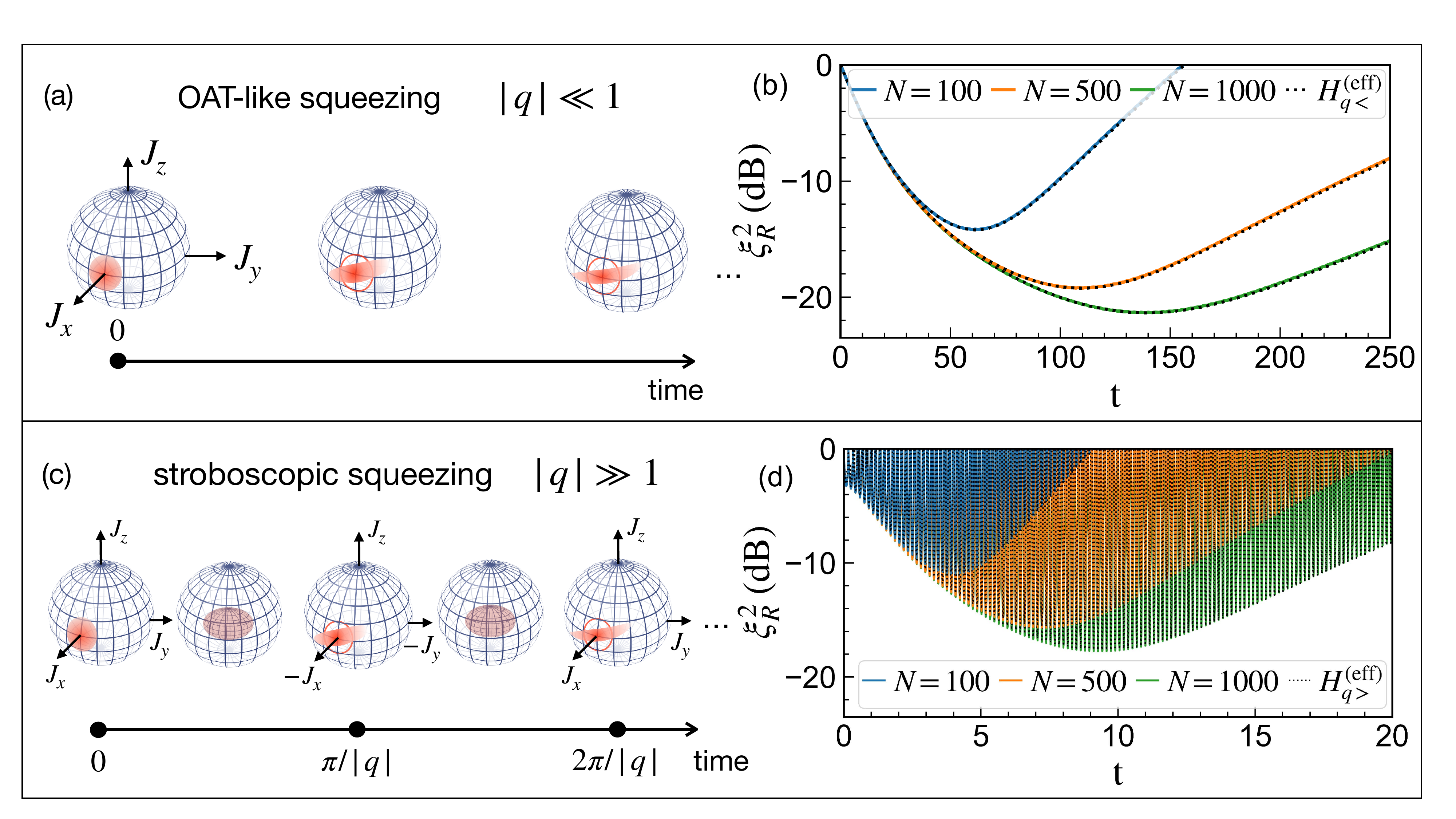}
\caption{{\bf Spin squeezing dynamics in spinor condensates.} (a) In the case $|q|\ll 1$ the system realizes an OAT-like dynamics; the state of the system is depicted schematically as an uncertainty region on the collective Bloch sphere for a spin of length $J = N$, although this length is not conserved for the model of Eq.~\eqref{e.H} and the state leaks into the volume of the Bloch sphere. (b) Exact data for the squeezing dynamics with $q = 0.1$, compared to the prediction of the effective Hamiltonian Eq.~\eqref{e.smallq}; (c) for  $|q|\gg1$ the spinor gas realizes instead a stroboscopic squeezing dynamics, in which OAT-like squeezing with a large collective spin occurs in rapid alternation with dynamics that makes the total spin depolarize and contract in length; (d) Exact data for the case $q = 20$, compared with the prediction of the effective Hamiltonian Eq.~\eqref{e.largeq}.}
\label{f.fig1}
\end{figure*}
\end{center}

\emph{Effective Hamiltonian for small $|q|$.} If  $|q| \ll 1$, the quadratic Zeeman term can be treated as a perturbation of the spin-dependent part of the Hamiltonian within degenerate perturbation theory. The latter reconstructs an effective Hamiltonian for the dynamics projected onto the Dicke subspace with maximal spin length. For $S=1$ the resulting effective Hamiltonian obtained by using the Schrieffer-Wolff transformation \cite{Hernandezetal2022,PhysRevB.109.214310} reads (see End Matter)
\begin{equation}
H^{(\rm eff)}_{q<} \approx \alpha_0 + \alpha_2 J_z^2   
\label{e.smallq}
\end{equation}
where, up to second order in perturbation theory, one has $\alpha_2  =  q~ (2 N+3)/[4 N(N+1)-3] +  2q^2 [1+2 N(N-1)] (2N+1)N/[(2 N-3) (2 N-1)^3 (2 N+1)] + O(q^3)$. 
This effective Hamiltonian has the form of the OAT (or planar rotor) Hamiltonian  $H_{\rm OAT} \propto J_z^2$~\cite{KitagawaU1993}, coming from the projection of the perturbation on the degenerate manifold. Further corrections lead to terms $\sim J_z^{2n}$ appearing at $n$-th order in perturbation theory (see End Matter). In particular the OAT dynamics starting from the CSS (sketched in Fig.~\ref{f.fig1}(a)) is known to lead to scalable squeezing, quantified by the squeezing parameter
\begin{equation}
\xi_R^2 = \frac{2NS~ \min_{\perp} {\rm Var}(J_\perp)}{\langle J_x \rangle^2}    
\end{equation}
where the minimization is taken over the collective spin components perpendicular to $J_x$. Spin squeezing is realized when $\xi_R^2 < 1$, and it corresponds to the net orientation of the collective spin becoming more sensitive to rotations than in any CSS. In particular squeezing witnesses the onset of  $(k+1)$-partite entanglement between the spins when $\xi_R^2 < 1/(kS+1)$ \cite{TrifaR2024}. 
%

\emph{Effective Hamiltonian for large $|q|$.} In the opposite limit $|q| \gg 1$, the fast one-body dynamics induced by the large quadratic Zeeman shift leads to a fast depolarization of the system on a time scale $t\sim 1/|q|$. One may conclude that the rapid vanishing of $\langle J_x \rangle$ (before correlations can be induced by the interaction term) prevents the system from developing scalable spin squeezing. And this behavior has indeed been observed theoretically for lattice spinor gases with dipolar interactions \cite{TrifaR2024} when the quadratic Zeeman shift exceeds significantly the strength of the dipolar coupling. 

Nonetheless the case of single-mode BECs, exhibiting all-to-all interactions between the atomic spins, turns out to be rather different. To understand how this is possible, a natural way of proceeding is to adopt the interaction picture of dynamics, namely to move to a ``rotating" frame evolving with the $q$-term of the Hamiltonian. 
This amounts to considering the state $|\psi_{\rm I}(t)\rangle = W_q(t)  |\psi(t)\rangle$, where $W_q(t) = e^{-iqN_0 t}$ and $|\psi(t)\rangle$ is the solution to the time-dependent Schr\"odinger's equation. This clearly defines a generalized rotating frame, since the quadratic Zeeman shift does not induce a conventional Larmor precession, but rather a nonlinear transformation of the individual spins for $S>1/2$. 

In the interaction picture for $S=1$ the state 
$|\psi_{\rm I}(t)\rangle$ evolves with the time-dependent Hamiltonian 
$H_{\rm I}(t) = 1/(2N) W_q(t) {\bm J}^2 W^\dagger_q(t) 
= H_{\rm diag} - H_{\rm off} (t)$ 
containing the diagonal time-independent term; and oscillating off-diagonal term
$H_{\rm off}(t)=1/N \left ( a_0^\dagger a_0^\dagger a_1 a_{-1} e^{-i2qt} + {\rm h.c.} \right )$.
The limit $|q| \gg 1$ implies the possibility of safely making a (generalized) rotating-wave approximation on $H_{\rm I}(t)$, i.e., of neglecting the terms rotating at the frequency $2|q|$.
As a result, when moving back to the lab frame the dynamics is essentially governed by the effective Hamiltonian
\begin{equation}
H^{(\rm eff)}_{q>} = \frac{1}{2N} \left [ J_z^2 - 2N_0^2 +  2N N_0 + N_0 + N \right ] - q N_0
\label{e.largeq}
\end{equation}
in which the originally SU(2)-invariant spin-dependent interaction term in (\ref{e.H}) has been reduced to an U(1)-invariant one. 
In particular, $H^{(\rm eff)}_{q>}$ crucially contains an OAT term, a nematic nonlinearity ($\sim N_0^2$), and effective detuning corrections linear in $N_0$.
%
The solution of the Schr\"odinger's equation takes the form 
$|\psi(t)\rangle \approx e^{-iH^{(\rm eff)}_{q>}t} |{\rm CSS}_x\rangle$, 
and it coincides with that in the ``rotating" frame $|\psi_{\rm I}(t)\rangle$ (up to a global phase) at times $t_p = \pi p /|q|$, with $p$ an integer. When $|q| \gg 1$ these times define a dense grid on the time scales $\sim O(1)$ of the dynamics governed by $H^{(\rm eff)}_{q>}$. Around the discrete times $t_p$, OAT-like squeezing dynamics is expected to appear in the system in a \emph{stroboscopic} fashion -- see Fig.~\ref{f.fig1}(d) for a sketch. 
 In fact, the dynamics  governed by the Hamiltonian $H^{(\rm eff)}_{q>}$  can be solved for analytically; the exact solution will be the subject of future work.   

\begin{center}
\begin{figure*}[ht!]
\includegraphics[width=1\textwidth]{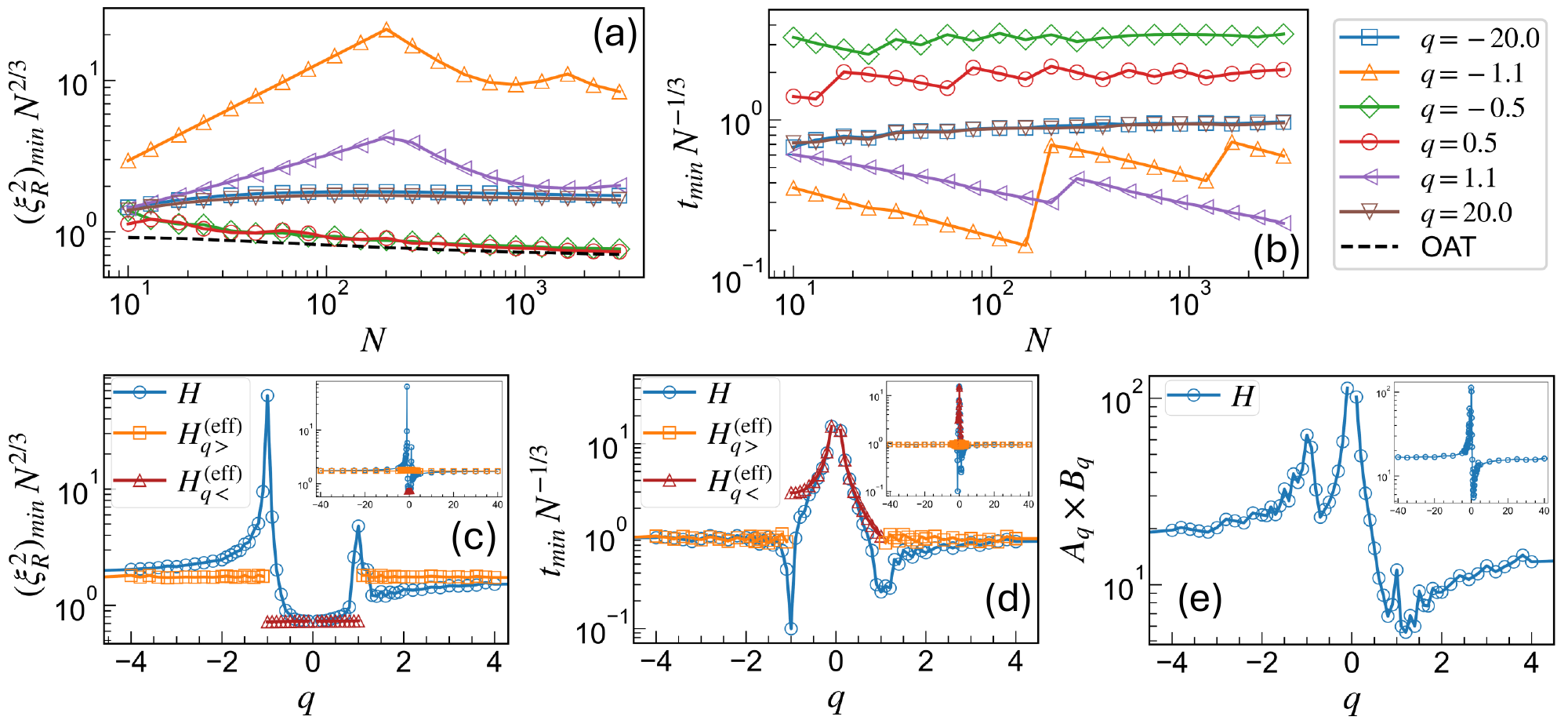}
\caption{{\bf Universal scaling properties.} 
%
(a)-(b) Scaling of the best squeezing $(\xi_R^2)_{\rm min}$ and the best squeezing time $t_{\rm min}$ with system size $N$ obtained from exact diagonalization, demonstrating the OAT scaling $(\xi_R^2)_{\rm min}\sim N^{-2/3}$ and $t_{\rm min}\sim N^{1/3}$ across the explored values of $N$. In panel (a), the dashed line indicates the scaling of optimal squeezing achieved with one-axis-twisting dynamics for $N$ spins $S=1$; (c)-(d) Dependence of the corresponding prefactors $A_q$ and $B_q$ on $q$, revealing strong variations of the best squeezing and the best squeezing time across parameter space.
(e) Optimal compromise between achievable squeezing and preparation time versus $q$, the lowest value is observed for $|q|\sim O(1)$.
}
\label{f.fig2}
\end{figure*}
\end{center}

\emph{Numerical results.} We quantitatively reconstruct the squeezing dynamics over the whole range of $q$ ratios via exact diagonalization (see End Matter). The original Hamiltonian in Eq.~\eqref{e.H} conserves the magnetization $J_z$, and it can therefore be efficiently diagonalized within each magnetization sector, which comprises only $O(N)$ states. Hence, we can obtain exact results for system sizes up to $N = 3 \times 10^3$. 

Fig.~\ref{f.fig1}(b)  shows that the effective Hamiltonian $H^{\rm (eff)}_{q<}$ of Eq.~\eqref{e.smallq} captures very accurately the full dynamics for $|q| \ll 1$, clearly exhibiting scalable squeezing.

In the opposite limit $|q| \gg 1$, scalability is preserved, modulo a fast modulation of the squeezing parameter at the frequency of $2q$ -- namely, scalable squeezing is realized stroboscopically. This modulation corresponds to an oscillation of the contrast $\langle J^x \rangle$ at the same frequency, as well as of the collective spin length $\langle {\bm J}^2 \rangle$, which shrinks when the contrast vanishes (see sketch in Fig.~\ref{f.fig1}(c)). Yet it remains of order $O(N^2)$, since the spin-dependent interaction energy $\langle {\bm J}^2 \rangle/(2N)$ remains of order $O(N)$. As seen in Fig.~\ref{f.fig1}(d), the effective Hamiltonian $H^{\rm (eff)}_{q>}$ captures the exact dynamics very accurately.

In the intermediate regimes $|q| \sim O(1)$, for which we do not have an effective theory, we systematically reconstruct the scaling properties of the squeezing dynamics, making use of the exact diagonalization. The scaling of the optimal squeezing $( \xi_R^2)_{\rm min}$ and of the optimal time $t_{\rm min}$ is shown in Fig.~\ref{f.fig2}(a-b): we observe that, for all the values of $q$ we explored \cite{footnote}, their scaling is consistent with that predicted for the OAT dynamics, namely $( \xi_R^2)_{{\rm min},N} \approx A_q N^{-2/3}$ and  $t_{{\rm min},N} \approx B_q N^{1/3}$ for $N \gg 1$. The cusp-like and jump-like features seen in Fig.~\ref{f.fig2}(a-b) are related to optimal squeezing switching between two successive minima of the $\xi_R^2$ oscillations induced by the $-qN_0$ term (see Fig.~\ref{f.fig1}(d); and see End Matter for further discussion).
Even though OAT scaling appears to be universal across the entire parameter space of the spinor-BEC Hamiltonian, the prefactors $A_q$ and $B_q$ of this scaling behavior display a strong dependence on $q$, which we estimate by calculating $A_q \approx N^{2/3} ( \xi_R^2)_{{\rm min},N}$ and  $B_q \approx N^{-1/3} t_{{\rm min},N}$ for $N=1000$. 
This dependence is shown in Fig.~\ref{f.fig2}(c-d). We observe that the squeezing parameter reaches the smallest values for $|q| \ll 1$; yet concomitantly, the time to achieve it diverges as $|1/q|$. Indeed, in the limit $q\to 0$ there is no dynamics, since the CSS becomes a Hamiltonian eigenstate. On the other hand, very strong squeezing is also achieved for $|q| \gg 1$, but over a time scale independent of $q$. The two limiting behaviors ($|q| \ll 1$ and $|q| \gg 1$) are not smoothly connected, as optimal squeezing displays sharp peaks centered around $|q|  = 1$. The best compromise between achieving the largest possible squeezing, and minimizing the time to achieve it, can be captured by the product $A_q B_q$, whose minimum, attained for $q\gtrsim 1$, signals an optimal strategy.

 \begin{center}
\begin{figure}[ht!]
\includegraphics[width=\columnwidth]{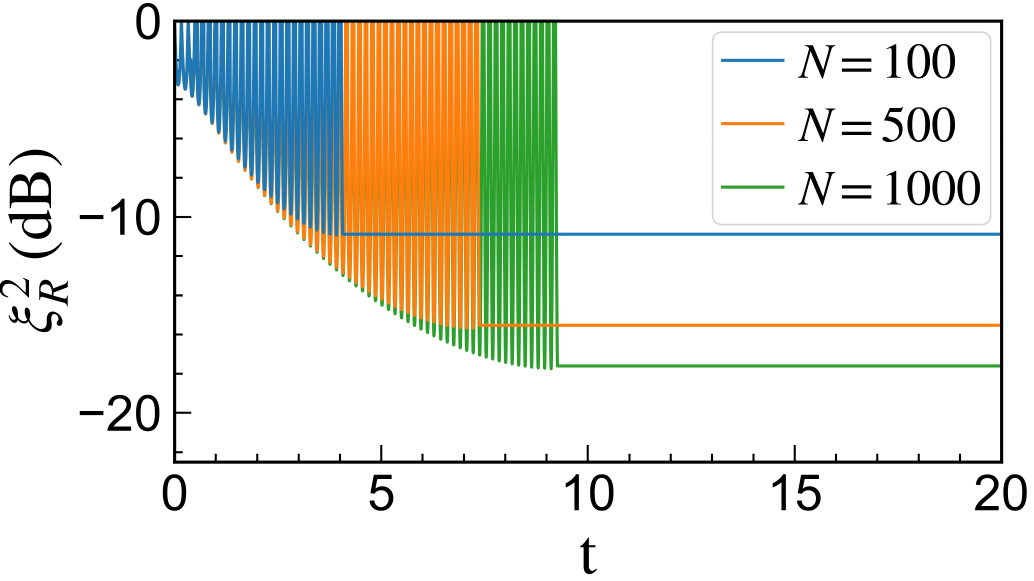}
\caption{{\bf Freezing of squeezed states.} If $q$ is turned off abruptly, the remaining Hamiltonian is SU(2) symmetric, so that the ensuing dynamics does not evolve the collective-spin properties, including squeezing. This allows to freeze the optimal squeezing -- here for  $q = 20$ (up to the optimal time $t_{\rm min}$) and increasing system size. }
\label{f.fig3}
\end{figure}
\end{center}

\emph{Freezing spin squeezed states.}
By switching $q$ off in the experiment, the remaining spin Hamiltonian becomes SU(2) invariant, $H_{q=0} = \bm J^2/(2N)$, and the corresponding evolution operator $U_{q=0}(t) = e^{-iH_{q=0} t}$ leaves all collective-spin operators unchanged. This opens the possibility of freezing the spin-squeezing dynamics during subsequent evolution. An illustration of freezing stroboscopic spin squeezing is shown in Fig.~\ref{f.fig3}. 
This feature is particularly convenient in a Ramsey sequence. There the atoms are exposed to an external field $B$ coupled to the (anti-squeezed) collective-spin component $J_\mu$ for a time $T$, namely they are subject to the unitary $U_B = e^{iJ_\mu B T}$. In the presence of a finite $q$, the interrogation time would be limited to $T \lesssim 1/|q|$, implying that fields $B \lesssim q$ might not be reliably estimated. However, the rotation $U_B$ commutes with the evolution operator $U_{q=0}$, so that, from the perspective of collective-spin observables, the interatomic interaction effectively disappears from the dynamics. Consequently, the interrogation time $T$ can in principle be arbitrarily long.

\emph{Conclusions.} In this work we have shown that the interplay between the spin-dependent interaction and the quadratic Zeeman shift in spinor Bose-Einstein condensates with $S=1$ leads to a universal entangling dynamics for the atomic spins, displaying the scaling properties of the paradigmatic OAT model. The quadratic Zeeman shift, albeit being a one-body term, is fundamentally responsible for breaking the SU(2) symmetry of the spin-changing collisions down to U(1), thereby allowing the Hamiltonian to evolve a (spin-separable) coherent spin state into spin-entangled states.   

We have specifically focused on spin squeezing in the early-time dynamics, for which scalability obeying OAT scaling can be achieved for any value of the Hamiltonian parameters. 
The central role of the quadratic Zeeman shift implies a high tunability of the entangling dynamics: indeed, the collective-spin dynamics can be entirely frozen by turning off this shift, so that the collective spin can be solely exposed to an external field to be sensed.  We would like to stress that these aspects are unparalleled in the case of ensembles of qubits.
Qubit Hamiltonians inducing scalable entanglement must possess interactions with at most U(1)-symmetry (i.e. XXZ interactions), such as OAT-like Hamiltonians \cite{KitagawaU1993,Perlin2020,Roscilde2021,Comparinetal2022,Blocketal2023}, or even a lower symmetry, such as two-axis-countertwisting or twist-and-turn Hamiltonians \cite{KitagawaU1993,Pezzeetal2018,Roscildeetal2025}. Their evolving effect on the collective spin cannot be turned off, unless their symmetry is raised to SU(2) e.g. via Floquet engineering \cite{Geieretal2021,Bornetetal2023}; or unless they are switched off altogether. 

Our quantitative predictions explicitly focus on the case of $S=1$ atoms, but the picture we present can be expected to hold for even larger spins carried by bosonic atoms such as $^{52}$Cr ($S=3$) \cite{Lepoutreetal2018, Lepoutreetal2019},  $^{168}$Er ($S=6$) \cite{Aikawaetal2012,Claudeetal2024} or $^{164}$Dy ($S=8$) \cite{Luetal2011} -- provided that the spin-dependent physics in the system is dominated by the $\sim {\bm J}^2$ interactions in Eq.~\eqref{e.H} (present for all magnetic atoms \cite{KAWAGUCHI2012253}), and the quadratic Zeeman shift. Atoms with $S>1$ can be prepared individually in a state which has a $>90\%$ overlap with the CSS$_x$ for $S=1$ individual atoms (see End Matter). For large $|q|$, the vast majority of atoms remains in the states with $m=0, \pm 1$, realizing therefore an effective $S=1$ system.  Our work paves the way for the realization of large-scale entangled states in large-$S$ spinor gases, and their practical exploitation for entanglement-assisted quantum sensing.

\emph{Acknowledgements.} We thank S. Nascimb\`ene for fruitful discussions. This work is supported by ANR (ANR-24-CE47-2670-02 'HighDy' project) and Polish National Science Centre SHENG project DEC-2023/48/Q/ST2/00087.



\bibliography{S1squeezing.bib}

\begin{thebibliography}{58}
\bibitem[{foo()}]{footnote}
\bibinfo{note}{At the exception of the case $q = -1$, for which we could not
  directly observe OAT scaling for the system sizes we had access to.}
\end{thebibliography}

\bigskip

\appendix
\centerline
{\bf END MATTER}

\bigskip

{\bf Effective Hamiltonian for $|q|\ll 1$.} The initial state of the evolution, $|{\rm CSS}_x\rangle$, belongs to the Dicke subspace of states with maximal ${\bm J^2} = N(N+1)$; yet the quadratic Zeeman shift leads to a dynamics leaking out of that subspace. Yet, if $|q|\ll 1$, the leakage can be considered as virtual, and, within perturbation theory, one can reconstruct an effective Hamiltonian governing the dynamics projected onto the Dicke subspace. 
The eigenstates of the unperturbed Hamiltonian $H_{q=0} = {\bm J}^2/(2N)$ are of the form $|J,M,\lambda\rangle$, namely joint eigenstates of ${\bm J}^2$ and $J_z$, with $\lambda$ a further quantum number (only required if $J<N$); their unperturbed eigenergies are $E_0(J) = J(J+1)/(2N)$.
Making use of the Schrieffer-Wolff transformation (SWT)  (see e.g. \cite{Hernandezetal2022}), the effective Hamiltonian up to second order in $q$ reads 
\begin{equation}
	{H}_{\rm eff} = {H}_0 + {H}^{(2)}_{\rm eff}~.
	\label{eq:effectivegen}
\end{equation}
Here
\begin{equation}
	{H}_0 = {\cal P}_{D} {H} {\cal P}_{D}
	\label{eq:H0diag}
\end{equation}
is the Hamiltonian projected onto the Dicke subspace,  
with ${\cal P}_{D}=|J=N,M\rangle\langle J=N, M|$ the corresponding projector. Moreover
\begin{equation}
	\label{eq:HSWgeneral}
{H}_{\rm eff}^{(2)} = {\cal P}_{D} ~{v}~ {G}_{N-2} ~{v}~
{\cal P}_{D} ,
\end{equation}
with  ${v}=-q {N}_0$ the perturbation, while 
\begin{equation}
	{G}_{N-2}= \sum_{M=-N+1}^{N+2} \sum_{\lambda}
	\frac{|N-2,M,\lambda\rangle\langle N-2, M,\lambda|}{E_0(N) - E_0(N-2)}
\end{equation}
is an operator which sums projectors onto the states $|N-2,M, \lambda\rangle$, weighted by the  energy mismatch denominator $E_0(N) - E_0(N-2)$, since the $v$ operator can connect Dicke states with $J=N$ only with states with $J=N-2$ \cite{Niezgoda_2019}. Using the matrix elements of bosonic operators $a_0, a_0^\dagger$ on the Dicke basis \cite{Niezgoda_2019} we obtain the following expressions: 
\begin{align}
    {H}_0&= \left ( \frac{1}{2 N} +  \frac{q(1+2N)}{3-4 N(N+1)} \right )  {\bm J}^2 \nonumber \\
     &-\frac{q(3+2 N)}{3-4 N(N+1)} ~{J}_z^2  - \frac{qN}{3-4 N(N+1)};
    \label{eq:effective}
\end{align}
\begin{align}
    &{H}^{(2)}_{\rm eff}
    = -2q^2\frac{(N-1)^2 N^2 (2N+1)}{(2 N-3) (2 N-1)^3 (2 N+1)} {N} \nonumber \\
    & + 2q^2 \frac{[1+2 N(N-1)] (2N+1)N}{(2 N-3) (2 N-1)^3 (2 N+1)} ~{J}_z^2 \nonumber\\
    & - 2 q^2 \frac{(2N+1)N}{(2 N-3) (2 N-1)^3 (2 N+1)}{J}_z^4.
    \label{eq:effective2}
\end{align}

In general, by inspecting the higher order terms, one can expect that the effective Hamiltonian takes the form of a polynomial of ${J}_z^2$, plus terms proportional to the square modulus of the total spin ${\bm J}^2$ and the number ${N}$,
\begin{equation}
	{H}_{\rm eff} = \sum_\sigma \alpha_\sigma ({J}_z^2)^\sigma + \beta_{J} {\bm J}^2 + \beta_{N} {N},
\end{equation}
where $\alpha_\sigma, \beta_{J(N)}$ are constants. Each order $\sigma$ of the perturbative analysis gives a term $\alpha_\sigma \propto q^\sigma$.


\bigskip

{\bf Exact diagonalization.} The time evolution of the spin-1 BEC, initialized in a coherent spin state and governed by the Hamiltonian in Eq.~\eqref{e.H}, is calculated via exact diagonalization (ED). The only off-diagonal term is of the form $a_{-1}^\dagger a_{+1}^\dagger a_0 a_0 + {\rm h.c.}$, which conserves the collective magnetization along the $z$ axis. Hence the Hamiltonian can be expressed in a block-diagonal form corresponding to the $2N+1$ magnetization sectors. 

The size of the basis can be further reduced thanks to the symmetry under rotation of $\pi$ around the $x$ axis, which flips the $J_z$ magnetization, of both the initial state and the Hamiltonian. Therefore the amplitude of basis states with finite positive magnetization $J_z$ is identical to those with opposite magnetization. Hence the simulation of the system requires only the time evolution of $N+1$ sectors (with $J_z \geq 0$), using diagonal Hamiltonian blocks  whose maximum  size is $(N/2 +1) \times (N/2+1)$ (in the $J_z=0$ sector for $N$ even). 

\begin{center}
\begin{figure}[ht!]
\includegraphics[width=0.9\columnwidth]{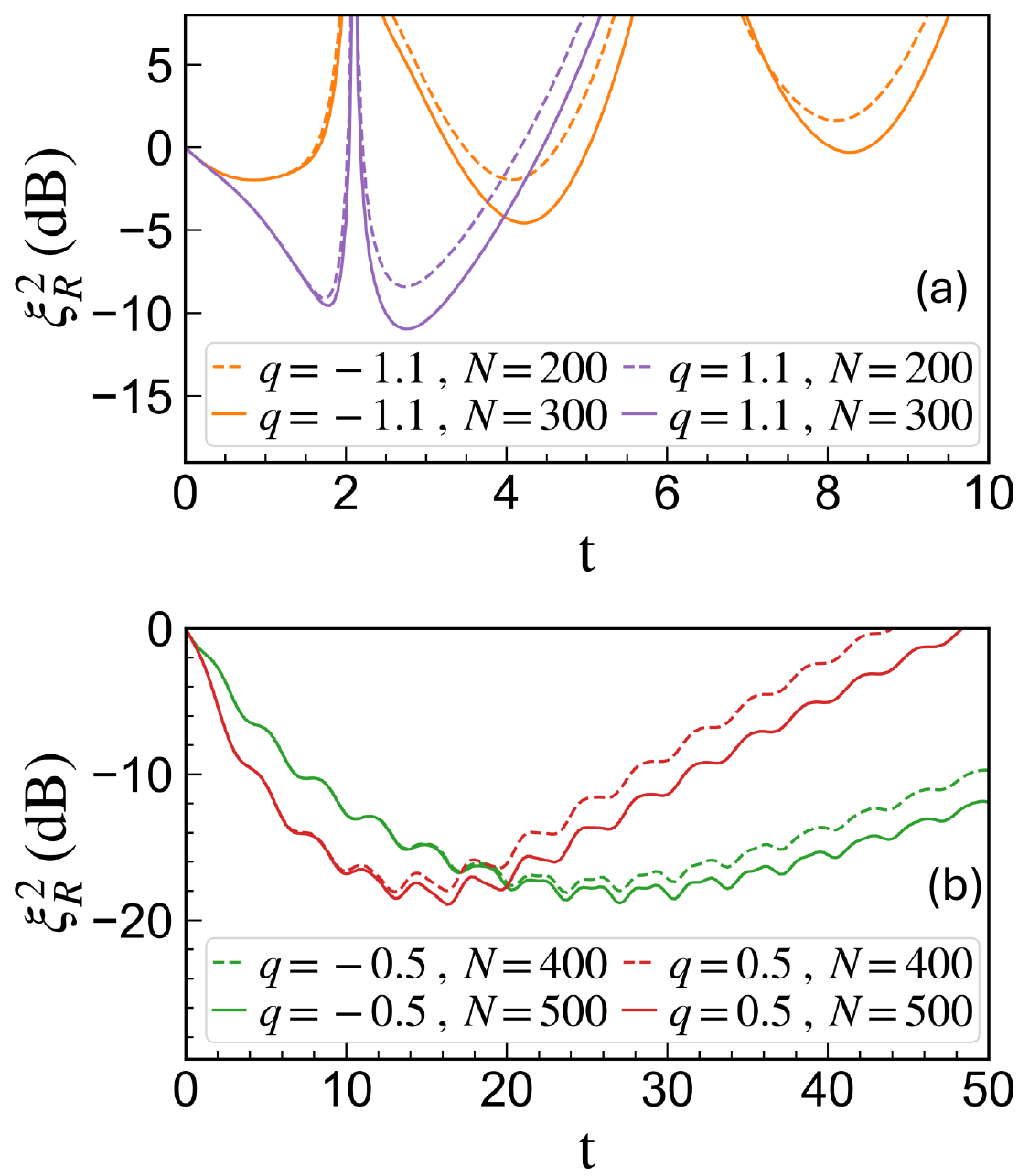}
\caption{{\bf Details on the scaling of optimal squeezing.}
Evolution of the spin squeezing parameter for (a) $q=\pm1.1$ and $N=200,300$ and (b) $q=\pm0.5$ and $N=400,500$. In both cases, upon increasing the system size the optimal squeezing is found to advance abruptly in time from one local minimum to the next in the oscillations of the squeezing parameter.}
\label{f.fig5}
\end{figure}
\end{center}

\bigskip

{\bf Scaling of optimal squeezing.} Fig.~\ref{f.fig2}(a-b) shows a seemingly non-monotonic scaling for both the optimal value and the corresponding time $t_{min}$ of squeezing parameter $\xi_R^2$, particularly so for $|q|\approx 1$. This non-monotonic behavior is not a breakdown of OAT scaling, but it can be understood by examining the system's dynamics explicitly. In the specific case of $q=\pm1.1$, the presence of a strong $q$ term in the Hamiltonian leads to large fluctuations of the squeezing parameter, with spin squeezing ($\xi_R^2<1$) appearing only stroboscopically over finite time windows separated by anti-squeezing ($\xi_R^2>1$). In particular,  as shown in Fig.~\ref{f.fig5}(a), the globally optimal value of the squeezing parameter over the entire time evolution can jump from a time window to the next upon increasing the size (from $N=200$ to $N=300$ in the figure), hence leading to a cusp in the optimal value and to a jump in the optimal time. For larger $q$ values the effect is much less pronounced, since squeezing time windows (of characteristic width $\sim q^{-1}$) form a much denser grid. For $q=\pm0.5$ (also shown in Fig.~\ref{f.fig2}(a-b)) we observe small-scale oscillations of the squeezing parameter on top of a smooth OAT-like behavior (see Fig.~\ref{f.fig5}(b)), to be attributed to leakage of the dynamics out of the Dicke subspace (i.e. physics beyond the SWT effective Hamiltonian described above). These oscillations result in fluctuations of the optimal time as seen in Fig.~\ref{f.fig2}(b).

 \begin{center}
\begin{figure}[]
\includegraphics[width=0.9\columnwidth]{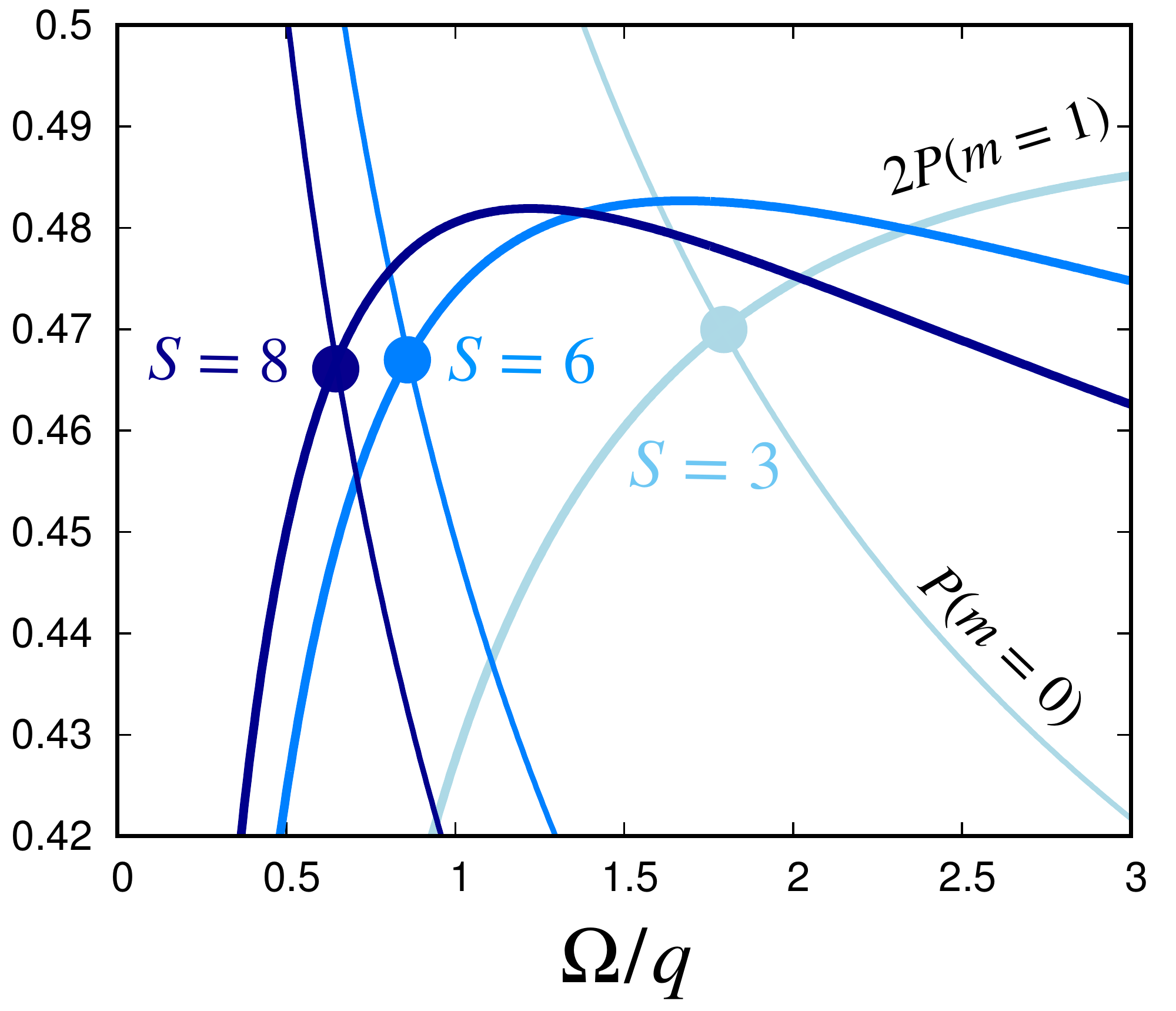}
\caption{{\bf Preparing a state close to the $S=1$ CSS$_x$ using $S>1$ atoms.} The probabilities $P(m=0)$ and $2 P(m=1)$ in the ground state of the Hamiltonian Eq.~\eqref{e.singleHam} cross at an $S$-dependent Rabi field $\Omega/q$. For this field, more that 90\% of the atomic population is concentrated on the states $m=0, \pm 1$, so that the system approximates an ensemble of effective $S=1$ atoms.}
\label{f.singlespin}
\end{figure}
\end{center}

\bigskip

{\bf Effective $S=1$ physics from $S>1$ atoms.} Atoms with $S>1$ can be prepared individually into a state with a large overlap with the CSS$_x$for $S=1$, by using the single-atom Hamiltonian \cite{Evrard2019} 
\begin{equation}
H = qS_z^2 - \Omega S_x
\label{e.singleHam}
\end{equation}
contaning the quadratic Zeeman shift $q$ (which we take here to be positive, $q>0$) and a Rabi field $\Omega$. Calling $P(m)$ the probability of an atom to be in the eigenstate $|m\rangle$ of $S^z$, the CSS$_x$ for $S=1$ has the characteristic that $P(m=0) = 2P(m=\pm 1) = 1/2$. The ground state of the Hamiltonian Eq.~\eqref{e.singleHam} for a single spin $S>1$ realizes the condition $P(m=0) = 2P(m=\pm 1)$ for a field $\Omega \sim q$, as shown in Fig.~\ref{f.singlespin}; and for that field one has $P(m=0) + P(m=1) + P(m=-1) \gtrsim 0.92$, namely the CSS$_x$ is realized with a large fidelity at the single-atom level, and only $< 10\%$ of the atoms
leak out of the $S=1$ manifold. 

If the ensuing dynamics is such that $|q| \gg 1$, the atomic populations will be prevented from further leaking into states with $|m|>1$, so that the behavior of the system can be expected to mimic  that of a system of effectively $S=1$ spins.

\end{document}